*Opinion paper*

# The Fallacy of Tumor Immunology

*Evolutionary pressures, viruses as nature's genetic engineering tools and T cell surveillance emergence for purging nascent selfish cells*

Tibor Bakacs[1], MD, DSc, Katalin Kristóf[2], MD, PhD, Jitendra Mehrishi[3], PhD (Cantab), FRCPath, Tamas Szabados[4], PhD, Csaba Kerepesi[5], PhD, Enikoe Regoes[6], PhD, and Gabor Tusnady[1], PhD, DSc

[1]Department of Probability, Alfred Renyi Institute of Mathematics, The Hungarian Academy of Sciences, 1053 Budapest, Reáltanoda str. 13-15. Hungary; [2]Department of Anesthesiology, Emergency and Intensive Care Medicine, University of Göttingen, Robert-Koch-Str. 40, Göttingen, 37075, Germany; [3]The Cambridge Stem Cell-Gene Therapy, Cultivated RBC Research Initiative [1]; [4]Department of Mathematics, Budapest University of Technology and Economics, Műegyetem rkp 3, Budapest, 1521, Hungary; [5]Department of Computer Science, Eötvös Loránd University, Pázmány Péter sétány 1/C, H-1117 Budapest, Hungary; [6]European Laboratory for Particle Physics (CERN) Geneva 23 CH-1211 CH-1211, Switzerland.

e-mail addresses: tiborbakacs@gmail.com (TB), katalin.kristof@med.uni-goettingen.de (KK), stemcell-crbc@virginmedia.com (JM), szabados@math.bme.hu (TSz), kerepesi@caesar.elte.hu (CsK), Enikoe.Regoes@cern.ch (ER), tusnady.gabor@renyi.mta.hu (GT).

*Running title: Viruses as nature's genetic engineering tools*

MS pages: 18; text words: 4727; text characters with spaces: 32293; Figure: 1.

"*It is not the strongest of the species that survives, nor the most intelligent, but the one most responsive to change.*" Charles Darwin

**Summary**

The US and Hungarian statistical records, before the dramatic medical advances, for the years 1900 and 1896, respectively, show 32% and 27% deaths attributable to infections, whereas only 5% and 2% due to cancer. These data can be interpreted to mean that (i) the immune system evolved for purging nascent selfish cells, which establish natural chimerism,

---

[1] Dr J N Mehrishi was formerly an Assistant Director of Research in the University of Cambridge, Department of Radiotherapeutics and a Research worker in the Department of Medicine. After leaving the University JNM for continuing studies and cooperation with international collaborators launched the Research Initiative (Independent of and separate from the University of Cambridge).

littering the soma and the germline by conspecific alien cells and (ii) defense against pathogens that represent xenogeneic aliens appeared later in evolution.
'Liberating' T cells from the semantic trap of immunity and the shackles of the 'two-signal' model of T cell activation, we point out theoretical grounds that the immune response to cancer is conceptually different from the immune response to infection. We argue for a one-signal model (with stochastic influences) as the explanation for T cell activation in preference to the widely accepted two-signal model for co-stimulation. Convincing evidence for our one-signal model emerged from the widespread adverse autoimmune events in 72% of advanced melanoma patients treated with the anti-CTLA-4 antibody (ipilimumab) that blocks an immune checkpoint. Harnessing the unleashed autoimmune power of T cells could be rewarding to defeat cancer. Assuming that immunization against isogeneic tumors also would be effective is a fallacy. (202 words)

1. **Introduction**

Alfred Tauber proposed that "…'immunity' may be a semantic trap that has confined our understanding of the immune system to only a narrow segment of defensive, aggressive functions" (1). Macfarlane Burnet suggested first regarding the evolutionary origin of adaptive immunity related to processes other than defense against pathogenic microorganisms (2). Satisfactory answers have not been available to explain 'why invertebrates including more than two million species in more than 20 phyla use only germline encoded innate immunity', or 'why vertebrates reject any allogeneic or xenogeneic transplanted tissue'. Rinkevich, then challenged the tacit assumptions and dogma that evolution of the immune system is pathogenically directed (3). He proposed that immunity developed as a surveillance operation to purge nascent selfish cells. Such cells could be isogeneic tumors from the host or transmissible allogeneic cells from kin organisms establishing natural chimerism that littered the soma and the germline by conspecific alien cells. According to Rinkevich the primary role of the vertebrate immune defense is to combat these parasitic events and preserve the individual homogeneity. Somatic compatibility systems that deter genotypes from being contaminated by maladapted alien genotypes might be the origin of immunity (4). Defense against pathogens, which are xenogeneic aliens appeared later in evolution.

Some current examples of parasitic allogeneic cells are instructive. Canine transmissible venereal tumor (CTVT) is a transmissible cancer allograft that rapidly spreads naturally in dogs worldwide (5). CTVT may have first arisen within a genetically isolated population of early dogs whose limited genetic diversity facilitated the escape of cancer from the immune surveillance system of the host. The Tasmanian devil facial tumor disease (DFTD) is another highly aggressive cancer allograft presenting a serious extinction risk for the Tasmanian devil population (6). DFTD arose in an island population with low genetic diversity. It seems that populations with limited genetic diversity may be particularly susceptible to the emergence and spread of transmissible cancers. This way, transmissible allogeneic tumors might have contributed to the evolutionary force shaping the class I immune surveillance system (7).

Based on the complementarity theorem of Dillon and Root-Bernstein (8) (9), we proposed that individual integrity can be preserved from parasitism with a limited repertoire. This is achieved by a homeostatic coupled system via internal dialogue between the positively selected, low affinity complementary T cells and host cells (10) (11). The role of regulatory T Cells (Foxp3+ Tregs) seems to be the closest analogy to the role of homeostatic T cells in our model, which is described in more details in (11).

The thrust of this paper is a fresh approach to reconsider the evolutionary role of viruses with respect to cancer over millions of years for adaptation and survival. The viruses are not just hostile invaders, but the molecular biologic tools of "*Nature*'s genetic engineering laboratory" that have been influencing and regulating key aspects of our biology (12).

Following Tauber, we liberate T cells from the semantic trap of immunity and suggest that the primary function of T cells is to prevent dedifferentiation that is, "…in a world in which necessity is represented by an inevitable disappearance of differentiation." [2] This way, T cells put strict limits on variations of host cells and prevent a natural tendency of people to develop tumors.

Unexpectedly, the first robust vindication for this proposal emerged from the widespread autoimmune adverse events in advanced melanoma patients receiving the checkpoint blocking anti-CTLA-4 antibody (ipilimumab) (13) as described below.

## 2. Ipilimumab clinical trials – Our alternative interpretation of severe, widespread autoimmune-related adverse events

The developers of the inhibitory anti-CTLA-4 antibody, ipilimumab, started with the premise that in an individual with no pathology other than cancer, most CTLA-4 expressing T-cells are either effector cells engaged in an anti-tumor response or regulatory T-cells actively opposing that response (14). Therefore, a CTLA-4 blockade would then selectively target T cells involved in the anti-tumor immune response. Although ipilimumab improved survival in a minority of metastatic melanoma patients, the vast majority suffered autoimmune-related adverse events (irAEs). A recent meta-analysis of ipilimumab mediated irAEs in 1265 patients from 22 clinical trials (15) included in the pooled analysis found a respective incidence of 72 % (95 % CI, 65–79; I2, 81.94) for all-grade irAEs and 24 % (95 % CI, 18–30; I2, 79.97) for high-grade irAEs leading to hospitalization or intravenous treatment. The incidence of all-grade irAEs varied according to the dosage of the drug, from 61 % (95 % CI, 56–66; I2, 0) in patients receiving ipilimumab at 3 mg/kg to 79 % (95 % CI, 69–89; I2, 85) in patients treated with ipilimumab 10 mg/kg. These results highlight the high risk of irAEs with anti-CTLA-4 drugs in patients with metastatic cancers. These are quite similar data to those of a retrospective review of safety data including 1498 patients treated with ipilimumab at various doses in 14 completed phase I–III trials, reporting inflammation drug-related adverse events in 64 %, with 18 % being of severe grades. [3] Life-threatening side effects, pathognomonic of acute graft-versus-host-disease (GVHD) and drug-related deaths (0.86%) have been reported in most trials (15) (16) (17). Notwithstanding, for some obscure reason, the CTLA-4 blockade is persistently thought to be tumor specific when clinical remission (partial or complete), or at least cancer stabilization, was noted for 60 % of patients who experienced an irAE. Objective tumor response rates were around 30 % in patients who developed autoimmune events, while 0–10 % of the other patients responded to treatment. In a study by Downey et al. (18), all complete responders experienced high-grade irAEs. These observations corroborate the idea of coupling autoimmunity and tumor immunity (15).

---

[2] Quoted from Norbert Wiener: *I Am* a *Mathematician*

We are inclined to suggest a cautious view that the widespread and dose-dependent irAEs of ipilimumab can be better explained by our one-signal T cell activation theory (11) (19) (20) (21) (22) than by the conventional two-signal T cell activation models. Our model suggests that all T cells are temporarily activated, expressing CTLA-4 that can then be targeted by anti-CTLA-4 antibodies. This is consistent with the *immunological homunculus* concept of Irun Cohen, who suggested that the immune system continuously responds to self (23) (24) (25) (26). Evolution may give an answer as to why such constant self peptide control is necessary.

3. **Viruses playing a pivotal role in evolution may increase the risk of DNA damage and cancer**

Viral genes outnumber cellular ones in the biosphere. The delivery of genes from virus to cell being invaded is far more overwhelming when compared with the reverse event, i.e. transfer of genes from host cells to viruses. Thus, viruses, apart from altering the cell surface molecular complex, the arena of specific interactions (27) (28) (29), have the unique ability to alter hundreds of genes with minimal genomic burden. Microarray analysis of transduced CD34+ cells with the GFP lentiviral vector revealed that a total of 513 genes were altered in terms of expression. Out of these, 183 (35.2%) were up-regulated more than twofold while 330 genes (64.4%) were down-regulated (30).

---

[3] Journal of Clinical Oncology, 2011 ASCO Annual Meeting Abstracts Part 1. Vol 29, No 15_suppl (May 20 Supplement), 2011: 8583

It had been proposed (12) that ancient viruses in evolution have spontaneously acted as essential editors of the host genome. This way, viruses may be thought of as the molecular biologic tools of "*Nature*'s genetic engineering laboratory" able to manipulate key aspects of our biology. Such natural genetic engineering in evolution will have contributed to the emergence of evolutionary innovations. A recent analysis confirmed that horizontal gene transfer is a hallmark of animal genome, although awareness of this virus-host ecology was not a part of the original Darwinian Theory (31) (32) (33) (34) (35) (36) (37).

Natural genetic engineering mediated by viruses is a double-edged sword. Whereas viral gene transfer speeded up the evolution of the species, viral remnants, e.g. jumping genes, represented a real danger to the individual by increasing the risk of DNA damage, cancer, and other maladies. Genomes with various interactions occurring are likely to be hotbeds of evolutionary conflict. [4]

Consistent with the hypothesis that defense against pathogens appeared later in evolution, epidemiological observations support the view that the immune system is far from being infallible against pathogens.

## 4. Before modern medicine, people succumbed to infections at much earlier age not living long enough to get cancer

The demographic transition from high to low mortality (38) occurred following the discovery of antibiotics and successful immunization programs. Before these medical achievements the likelihood of an individual dying prematurely from infectious diseases was as high as 40%. [5] As Mukherjee emphasized, prior to the miracles of modern medicine, "people didn't live long enough to get cancer. Men and women were long consumed by tuberculosis, dropsy, cholera, smallpox, leprosy, plague, or pneumonia (39)."

In contrast to the 40% death rate by infections, only one-third of humans are struck by cancer, mainly with advancing age (40). Fortuitously, good supporting historical evidence is available in the Statistical Yearbook of Hungary from 1896 about all-causes and cause-specific mortality. We created an interactive figure using the visualization tool Krona (41). [6] [7] The data show that deaths due to infections were 27%, whereas deaths due to cancer were only 2%. It must be noted that the mortality rate of 27% from infectious diseases is a conservative estimate, since pneumonia, bronchitis, meningitis and encephalitis were not included therein in the infectious disease category. It is noteworthy that a similar difference between the mortality rate from infectious disease and other disease states was recorded in the USA. Cutler and Meara reported that at the beginning of the 20th century, death due to infections was 32%, whereas only 5% due to cancer. [8] In the low-income countries, where the miracles of modern medicine are still not readily available, this ratio (28% vs. 6%) had not changed much by 2012. [9] [10] [11]

---

[4] http://www.the-scientist.com//?articles.view/articleNo/42274/title/Wrangling-Retrotransposons/
[5] http://www.cdc.gov/mmwr/preview/mmwrhtml/mm4829a1.htm#fig1
[6] http://digitalia.lib.pte.hu/books/magyar-statisztikai-evkonyv/htm/1896/htm/094.htm
[7] http://kerepesi.web.elte.hu/causes_of_death/1896_leading_causes_of_death_en.txt-krona.html

### 5. Slow growing tumor cells induce tolerance

Carcinogenesis is a long-lasting step-by-step progression of early-stage lesions of cancer into frankly malignant cells. In addition, it is noteworthy that it takes as long as 12 years for cancer cells to reach a population size of $10^9$ cells contained within ~0.5 cm$^3$, weighing ~0.5 g (42). Consistent with this, the average age of women with pre-invasive lesions was about 20 years lower than for those with invasive lesions. [12] The risk of cancer increases exponentially with age (43). The risk of breast cancer, for example, increases from 1 in 400 at thirty years of age to 1 in 9 at seventy years of age. Age-incidence curves rise sharply above the age of 50 years and are informative about the dynamics of tumor progression, the straight line showing a fit with power 4.8 (44).

The slow growth of tumor cells is consistent with Pardoll's suggestion that specific immune surveillance systems operate at early stages of tumorigenesis, whereas established tumors induce immune tolerance (45). The latter phenomenon is explained by the discontinuity theory of immunity claiming that the speed of antigenic change determines T cell activation. That is, the elimination of target cells is induced by an antigenic discontinuity, following a sudden modification of molecular motifs with which T cells interact (46).

The paradox of cancer appears not to be *"why does it occur", but rather "why does it occur so infrequently"* (47). It perhaps bears repeating that in fact, most human malignant tumors are latent for many years and became 'old' by the time they are detectable clinically, when termed incipient cancer. Although two out of three humans never develop clinically detectable cancer (40), most individuals with no apparent pathology, but having died of trauma, at autopsies were discovered to have been harboring unsuspected microscopic primary cancers (48) (49). The risk of suffering any cancer before the age of 40 is ~2%, but by age 80 this risk increases to 50% (50). For this so called tumor dormancy, it was suggested that cancer may be thought of as a chronic disease, which is kept in check by the patients' own immune system and physiological mechanisms (51).

---

[8] See Table 3 in http://www.nber.org/papers/w8556
[9] http://apps.who.int/gho/data/view.main.CODWBINCLOINCV?lang=en
[10] http://kerepesi.web.elte.hu/causes_of_death/low_income_leading_causes_of_death.txt-Krona.html
[11] http://kerepesi.web.elte.hu/causes_of_death/high_income_leading_causes_of_death.txt-Krona.html
[12] Siddhartha Mukherjee: *The Emperor of All Maladies*, p.290

## 6. The explosive replication speed of microbes can only be controlled with non-cytopathic mechanisms

Numbers of bacteria double in 20 min, whereas viruses produce more than 1,000 progeny in a few hours generating hundred-times more virus infected cells within a few days than cancer cells develop during twelve years (see below). The question arises as to how do specific cytotoxic T lymphocytes (CTL) cope with virus infected cells?

In this connexion hepatitis B and C virus (HBV, HCV) infections are good examples. According to Guidotti and Chisari (52), in the unlikely event of all the $10^8$ HBV-specific CTL in the entire body entering the liver at the same time and all the $10^{11}$ hepatocytes quite commonly infected, for every 1,000 infected hepatocytes, there would be only one specific CTL in the liver to cope with the infection. Obviously, 1:1000 ratio would be totally inadequate for cytotoxic mechanism alone. Nevertheless, the immune system of most infected patients clears the virus within a few weeks without serious liver disease. This fact indicates the contribution of non-cytopathic mechanisms. Similarly, this occurs in HCV infections as well (53).

## 7. The law of independent T cell activation is consistent with recent clinical observations

The consensus view still is that the immune system is a complex and powerful defense mechanism (54). In order to keep this power under control T cell antigen receptor (TCR) input must be complemented by CD28 co-stimulation to promote interleukin-2 (IL-2)–dependent proliferation, as described by the classic "two-signal" model of T cell activation (55). This is taken to mean that each cell requires the conjoint signals within it for these two receptors triggering activation to a state suitable for cell division (56).

In contrast, the law of independence of T cell activation described by Gett and Hodgkin (57) states that the internal mechanisms that control the rate of division, the likelihood of surviving and the likelihood of undergoing a differentiation operate independently within a cell. In fact, the strength of a T cell response can be predicted by adding together the underlying signal components from the TCR, co-stimulatory receptors, and cytokines. This law resolved the co-stimulation paradox and provided a quantitative paradigm for therapeutically manipulating immune response strength (58). Consequently, there is no need for an obligatory co-stimulus for the decisions between tolerance and activation.

This law is confirmed by recent clinical observations, which demonstrated that one signal alone was sufficient to trigger uncontrolled T cell stimulation via the CD28 receptor or the CTLA-4 receptor (19). The cytokine storm induced by the "superagonist" anti-CD28 mab (TGN1412) in the Northwick Park, Harrow, clinical trial catastrophe demonstrated that T cells, depending on the high ligand concentration (see in (59) especially fig. 2), can be activated via a single receptor. Similarly, anti-CTLA-4 antibodies (during ipilimumab therapy) were able to stimulate the T cell system indirectly by blocking the CD28 antagonist CTLA-4 co-receptors. In these interactions T cell pathways responsible for immune down-regulation were interrupted resulting in a dose-dependent, unrestrained, pan-lymphocytic T cell activation. This then turned homeostasis into overt autoimmunity (60) (61), thereby provoking an autologous graft versus host-like disease (GVHD) with severe, life-threatening autoimmune side effects.

## 8. The fallacy of the infectious disease vaccination model of cancer immune therapy

One of the greatest triumphs of medicine was the discovery of immunization against infectious diseases. Successful vaccination programs have been developed against 27 different diseases. Vaccination against smallpox, which killed 300 to 500 million people even in the 20th century, enabled the infection to be declared eradicated from the world in 1980. [13]

Following the success of vaccines against xenogeneic infectious diseases, the tacit assumption, historically, has been that host immunity should be protective against isogeneic cancer as well. Following the simple principle of logic, assuming that immunity should be successful to defeat cancer as well, we submit that this makes it a fallacy. It is useful to consider the following reasons: (i) most spontaneous tumors in humans have no neoantigen (they are part of self), (ii) each tumor is a unique entity – a unique invention of nature (62), (iii) consistent with this, nearly all neoantigens in ipilimumab treated melanoma patients were patient-specific and most likely represent "passenger" mutations that do not directly contribute to tumorigenesis (63), but may have enhanced ipilimumab induced GVHD and (iv) tumors grow very slowly compared to pathogens (even a fast growing tumor with a doubling time of 10 days, less than 10 tumor cells will be present in a month, whereas $10^{11}$ hepatocytes are infected during the same time; see above). Due to these characteristics human tumors are either unable to activate a sufficient number of antigen presenting cells (APCs) or APCs are less efficient in responding to isogeneic variants to promote interleukin-2 (IL-2)–dependent co-stimulation of T cell before tolerance will have generated (45).

## 9. Tolerance breakdown is required for eradication of isogeneic tumors

Conventional cancer immunotherapy trials conducted with the best available science resulted in anecdotal responses. It is generally acknowledged that immunotherapy of cancer has not quite fulfilled the great promise and hopes of conquering cancer losing credibility (64).

---

[13] https://www.gov.uk/government/publications/immunity-and-how-vaccines-work-the-green-book-chapter-1

Studies initiated by James P. Allison led to the breakthrough to treat a variety of malignancies. This was achieved by a prolonged overstimulation via immune checkpoint blockade by antibodies that target negative regulators of T-cell activity such as the cytotoxic T lymphocyte-associated antigen 4 (CTLA-4) and the programmed cell death protein 1 pathway (PD-1/PD-L1) (65). The checkpoint blockade extends the expiration of activated T-cell at various stages of the immune response.

CTLA-4 receptors of T cells are an indispensable braking mechanism on T cell activation to ensure tolerance to self-tissues. Should CTLA-4 not function due to a genetic deficiency or if it is blocked by various manipulations, CD28 functions unopposed and swings the balance in favor of immune stimulation resulting in breakdown of tolerance (19). Anti-CTLA-4 antibody (ipilimumab) improved survival of metastatic melanoma patients by disabling the brakes of T cells. Thus, the price we pay for reversing immunosuppression in cancer by a prolonged immune checkpoint blockade is the generation of uncontrolled T-cell activation.

Successful targeting CTLA-4 has created enthusiasm for clinical approaches targeting other immunologic checkpoints, namely PD-1/PD-L1. After years of skepticism on lung cancer immune therapy in patients with metastases, trials on anti-PD-1/PD-L1 antibodies (nivolumab and pembrolizumab, respectively) provided results never observed with previously known drug categories (66).

## 10. Worldwide increasing incidence of asthma affecting over 300 million people an instructive example of immune imbalance provoked by one-signal

The anti-CD28 (TGN1412) and anti-CTLA-4 (ipilimumab) antibody trials demonstrated that one-signal alone was sufficient to activate the immune system (see above). Another instructive example of immune imbalance provoked by one-signal is the rising prevalence of asthma.

Especially, from after the 1960s in the so-called "Westernized" countries of the world, rapid urbanization and industrialization have increased air pollution. The exposures of populations have resulted in a sharp increase in the prevalence, morbidity, and mortality of asthma. Air pollutants (e.g. particulate matter) induce expression of co-stimulators, such as B7 (CD80/86) molecules, on the antigen presenting cells (APCs) providing prolonged "signal 2" that then initiate T cell responses. Thus, environmental changes appear to be directly responsible for the pathophysiology of asthma, driving the development of the T helper type 2 ($T_h2$)-biased immune responses and the overproduction of cytokines. Under such conditions then the immune system gets dysregulated mistaking pollen for pathogen reacts when it really should not. The prevalence of asthma is rising also in low and middle income countries as they adopt a more Western-type lifestyle. In China, for example, where 1.9 billion tons of coal is burnt each year to meet the 70 to 75% of the energy needs, outdoor pollution is associated with more than 300,000 deaths and 20 million cases of respiratory illnesses annually. So far asthma has affected more than 300 million people worldwide and the incidence is rising (67) (68).

The consequences of prolonged overstimulation of the B-7/CD28 axis by drugs or pollutants are persuasive enough to consider a critical re-examination of the conventional two-signal theory, invoking an obligatory co-stimulus for T cell activation as described below.

**THE ONE-SIGNAL MODEL – A PAINTING WITH BROAD BRUSH STROKES**

Numerous receptor-mediated signals are delivered to T cells, governing their survival, differentiation and proliferation. For the sake of simplicity, only two positive and one negative signals are depicted in Figure 1.

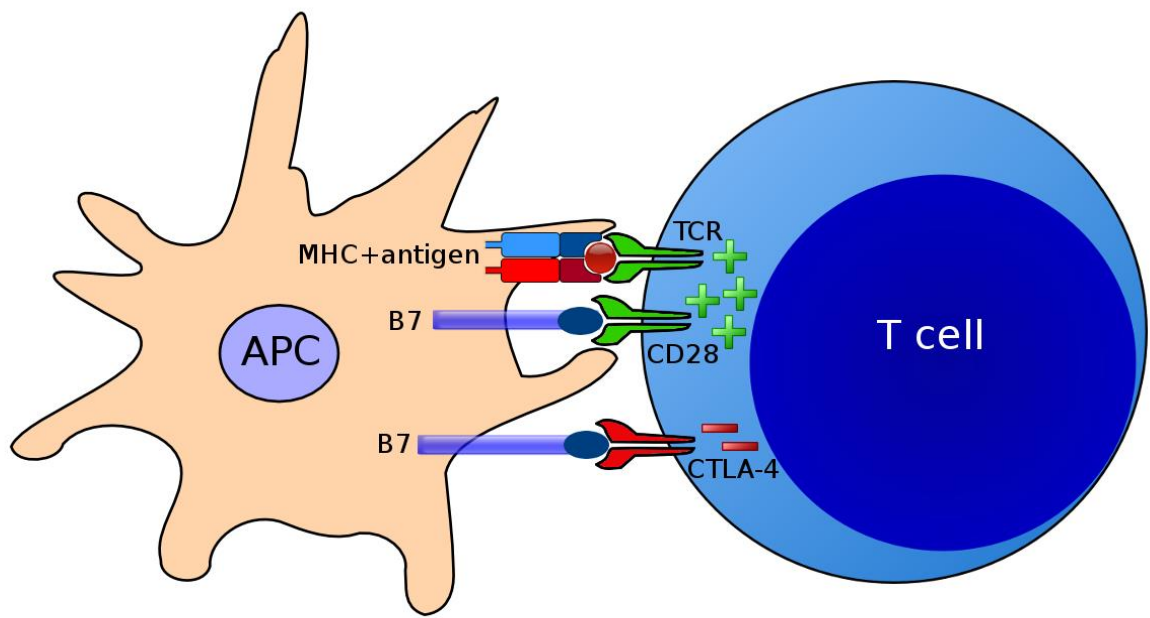

**Figure 1.** *Activation of T-cells*: Numerous receptor-mediated signals are delivered to T cells, which direct their survival, activation, differentiation and proliferation. For the sake of simplicity, here only an APC and a T cell are depicted with two ligands on APC (MHC and B7 [CD80/CD86]) and three receptors on T cell (TCR; CD28; and CTLA-4). The TCR and the CD28 receptor mediated signals are stimulatory, whereas the CTLA-4 receptor mediated signal is inhibitory in our model. Further details see in the text.

In our model stochastic processes govern all the events. Based on the law of independent T cell activation (57), the likelihood of survival, activation, undergoing a differentiation change and clonal division operate independently within a cell. In other words, one signal should be sufficient to instigate these events. Signal strength (via one or several receptors) determines the outcome of T cell activation. Low affinity, short-lasting TCR ligation ensures T cell survival, increased ligation time and or affinity induces cytotoxicity, whereas the strongest signal (long ligation time and or high affinity) induce clonal division. This may require stimulation via more than one receptor (e.g. TCR and CD28) such that the TCR-dependent effects are strengthened by co-stimulation amplifying the T cell number exponentially through minimal kinetic alterations. The activities of the model are described in greater detail hereunder the following headings: (1) T cell survival (2) tumor prevention; (3) defense against primary infections; (4) secondary immune response; (5) autoimmune disease and (6) iatrogenic tolerance breakdown.

1. *T cell survival*

As already stated, the homeostatic coupled system functions via an internal dialogue between positively selected low affinity complementary T cells and host cells (10) (11). Recognition of ubiquitous and constitutive self-antigens by complementary T cells not only reliably sustains natural tolerance preventing dedifferentiation, but also ensures attacking cells presenting non-self peptides (see below).

2. *Tumor prevention (surveillance)*

The success of multicellularity depends upon the evolution of mechanisms that are able to suppress the ability of virtually every cell in an organism with the information and the potential to propagate rapidly (50). Rinkevich suggested that the immune system has developed as a surveillance machinery for nascent selfish cells stemming from a kin organism or from transformed cells within the organism of origin (3). Protective mechanisms that evolved over millions of years are indeed capable to keep the incidence of cancer very low (~2%) during reproductive age (50). It should, however, be noted that despite appearances, the mechanism of primary protection against cancer is different from primary protection against infections (see below).

We hypothesize that cancer protection is carried out via cognate (complementary) TCR-MHC interactions (see Fig. 1) such that T cells keep the number of somatic cells constant. Paraphrasing Georg Klein (40), it would appear that evolution may have exploited over expression of a relatively limited number of common resistance genes to nip in the bud the incipient cancerous foci. Such preventive protection is all the more important since cancer is a state in which the epigenome is allowed to have greater plasticity than it is supposed to have in normal somatic tissues. It was argued that this increased epigenetic plasticity allows for selection in response to the cellular environment for cellular growth advantage at the expense of the host (69).

Up-regulated genes in transformed cells increase the expression of self peptides, which in turn, increase their affinity for interaction with the TCRs. This way, short-lasting life-sustaining physiological stimulation of T cells is extended into a longer-lasting one that induces cytotoxicity. This local destructive autoimmunity eliminates altered (pre-cancerous) host cells. This is encompassed in the real meaning that molecular complementarity between TCR and MHC molecules puts strict limits on variations. The proof-of-principle that amplified or overexpressed genes are capable of inducing robust antitumor efficacy in T cells without destruction of normal tissues was recently demonstrated (70).

Notwithstanding, cancer is virtually inevitable in complex, long-lived, multicellular organisms. Extrapolating from the risk of affliction with any cancer, practically everybody will have developed cancer as human lifetime approaches one-hundred years (see figure 4 in (50)). This is due to the fact that somatic mutations inevitably accumulate with time and capable to overcome the suppressive mechanisms.

3. *Defense against primary infections*

The probability is greater that the presentation of foreign peptides decreases rather than increasing the affinity for interaction with the TCRs during primary infections (10). While specific TCR-MHC contacts are inhibited, CD80/86-CD28 engagements are not. Microbial products (e.g. endotoxin) increase the local concentration of the CD80 and CD86 ligands on APCs (e.g. dendritic cells) by stimulating toll like receptors (TLRs). CD28 receptors of bystander T helper cells ($T_h$) will then be saturated with CD80/86 ligands. Consequently, $T_h$ cells in the anatomical region are activated via the CD28 receptor alone (59). This triggers a limited beneficial local cytokine storm unleashing polyclonal T cytotoxic ($T_c$) cell proliferation to attack infected cells. These events initiate predominately immunopathology and to a lesser extent, autoimmunity by inducing indiscriminately anti-self and anti-non-self killing. This may well be thought of as a physiological local transplantation reaction.

4. *Pathogen specific secondary immune response*

In the presence of ongoing robust non-associative CD80/86-CD28 interactions (co-stimulation), there is always a possibility that rare T cell (and B cell) clones with higher affinity may well recognize foreign peptides (antigens) via MHC-Ag-peptide-TCR signal (or via BCR), particularly when a significant fraction of host cells is infected and viral load is high (for example in hepatitis, see in (53)). Such higher affinity specific interactions would then drive activation, proliferation of T cell clones, and eventually lysis of infected cells, as described by the conventional two-signal models. Having cleared the infection, specific T cells would expand into memory type T cell clone, while B cells would differentiate into antibody producing plasma cells (71). Specific T and B cell activation, proliferation and lysis of infected cells, therefore, obey the rules of the conventional two-signal model.

   5. *Accidental autoimmune disease*

During an infection, when infected host cells lose their complementary $T_c$ cell contact, autoreactive $T_c$ cells with high affinity for a self peptide-MHC complex may be generated randomly, albeit with a small probability. This is consistent with the observations that autoimmunity might be thought of as a by-product of the immune response to microbial infection (72).

   6. *Ipilimumab therapy and unanticipated consequences of iatrogenic breakdown of tolerance*

Ipilimumab therapy that induces blockade of CTLA-4 disabling the brakes on T cells, artificially prolongs the survival signal of complementary (autoreactive) T cells and turns homeostasis into overt autoimmunity. This induces a dose-dependent, unrestrained, pan-lymphocytic T cell activation, which lasts so long as the CTLA-4 receptor blockade is sustained. This unfortunately results in unintended severe widespread, often life-threatening autoimmune side effects, including autologous graft versus host disease (GVHD). Under such conditions a genuine monoclonal autoimmunity may also be triggered (19) (20) (21).

**Conclusions**

Immunotherapy trials conducted with the best available science so far have not quite fulfilled the great promise and hopes of conquering cancer. Most spontaneous tumors are a part of self, a unique invention of nature. Nearly all neoantigens represent "passenger" mutations that do not directly contribute to tumorigenesis. The autoimmune power of T cells unleashed by the blockade of immune checkpoints should be harnessed for curing cancer. Tacitly assuming that immunization, which was so dramatically successful overcoming xenogeneic infections, should also be successful against isogenic tumours to defeat cancer surely is a fallacy.

While we pointed out the pitfalls and drawbacks of ipilimumab, we also recognized its potentials that had remained undiscussed. We suggested extending the use of ipilimumab to eradicate minimal residual disease (MRD) following induction of complete remission by high dose chemotherapy and autologous stem cell transplantation, or following reduced intensity conditioning in preparation for allogeneic stem cell transplantation (SCT) using donor lymphocyte infusion (DLI) if indicated (21). On the other hand, in patients with high tumor burden, the forces of the immune system liberated by the co-stimulatory anti-CTLA-4 antibody blockade could be tethered to the tumor cells without collateral damage to normal cells using pretargeting (20).

**Acknowledgements**

The authors thank Professor Alfred I. Tauber, Chairman of the Board of Governors, University of Haifa, Israel and Baruch Rinkevich, senior scientist, Israel Oceanographic and Limnological Research (IOLR), Haifa, Israel for their critical reading, comments and suggestions for the MS.

**Conflict of interest**

T.B. has a pending patent application and he is the CSO of Pret Therapeutics Inc.

Reference List

(1) Tauber AI. A tale of two immunologies. In: Grignolio A, editor. Immunology Today: Three Historical Perspectives under Three Theoretical Horizons.Bologna: Bononia University Press; 2010. p. 15-34.

(2) Burnet M. The Clonal Selection Theory of Acquired Immunity. Cambridge: Cambridge University Press; 1959.

(3) Rinkevich B. Primitive immune systems: are your ways my ways? Immunol Rev 2004 April;198:25-35.

(4) de Boer RJ. The evolution of polymorphic compatibility molecules. Molecular Biology and Evolution 1995;12(3):494-502.

(5) Murchison EP, Wedge DC, Alexandrov LB, Fu B, Martincorena I, Ning Z et al. Transmissable dog cancer genome reveals the origin and history of an ancient cell lineage. Science 2014 January 24;343(6169):437-40.

(6) O'Neill ID. Tasmanian devil facial tumor disease: insights into reduced tumor surveillance from an unusual malignancy. Int J Cancer 2010 October 1;127(7):1637-42.

(7) Hickman HD, Yewdell JW. Mining the plasma immunopeptidome for cancer peptides as biomarkers and beyond. PNAS 2010 November 2;107(44):18747-8.

(8) Root-Bernstein RS, Dillon PF. Molecular complementarity I: the complementarity theory of the origin and evolution of life. J Theor Biol 1997 October 21;188(4):447-79.

(9) Dillon PF, Root-Bernstein RS. Molecular complementarity II: energetic and vectorial basis of biological homeostasis and its implications for death. J Theor Biol 1997 October 21;188(4):481-93.

(10) Bakacs T, Mehrishi JN, Szabados T, Varga L, Szabo M, Tusnady G. T Cells Survey the Stability of the Self: A Testable Hypothesis on the Homeostatic Role of TCR-MHC Interactions. Int Arch Allergy Immunol 2007 May 30;144(2):171-82.

(11) Szabados T, Bakacs T. Sufficient to recognize self to attack non-self: Blueprint for a one-signal T cell model. Journal of Biological Systems 2011;19(2):299-317.

(12) Witzany G. A perspective on natural genetic engineering and natural genome editing. Introduction. Ann N Y Acad Sci 2009 October;1178:1-5.

(13) Hodi FS, O'Day SJ, McDermott DF, Weber RW, Sosman JA, Haanen JB et al. Improved survival with ipilimumab in patients with metastatic melanoma. N Engl J Med 2010 August 19;363(8):711-23.

(14) Curran MA, Callahan MK, Subudhi SK, Allison JP. Response to "Ipilimumab (Yervoy) and the TGN1412 catastrophe". Immunobiology 2012 June;217(6):590-2.

(15) Bertrand A, Kostine M, Barnetche T, Truchetet ME, Schaeverbeke T. Immune related adverse events associated with anti-CTLA-4 antibodies: systematic review and meta-analysis. BMC Med 2015;13:211.

(16) Graziani G, Tentori L, Navarra P. Ipilimumab: A novel immunostimulatory monoclonal antibody for the treatment of cancer. Pharmacological Research 2012 January;65(1):9-22.

(17) Voskens CJ, Goldinger SM, Loquai C, Robert C, Kaehler KC, Berking C et al. The price of tumor control: an analysis of rare side effects of anti-CTLA-4 therapy in metastatic melanoma from the ipilimumab network. PLoS ONE 2013;8(1):e53745.

(18) Downey SG, Klapper JA, Smith FO, Yang JC, Sherry RM, Royal RE et al. Prognostic factors related to clinical response in patients with metastatic melanoma treated by CTL-associated antigen-4 blockade. Clin Cancer Res 2007 November 15;13(22 Pt 1):6681-8.

(19) Bakacs T, Mehrishi JN, Moss RW. Ipilimumab (Yervoy) and the TGN1412 catastrophe. Immunobiology 2012 June;217(6):583-9.

(20) Bakacs T, Mehrishi JN, Szabo M, Moss RW. Interesting possibilities to improve the safety and efficacy of ipilimumab (Yervoy). Pharmacol Res 2012 August;66(2):192-7.

(21) Slavin S, Moss RW, Bakacs T. Control of minimal residual cancer by low dose ipilimumab activating autologous anti-tumor immunity. Pharmacol Res 2013 November 4.

(22) Bakacs T, Mehrishi JN. Anti-CTLA-4 therapy may have mechanisms similar to those occurring in inherited human CTLA4 haploinsufficiency. Immunobiology 2014 December 6;220:624-5.

(23) Cohen IR. The cognitive paradigm and the immunological homunculus. Immunol Today 1992 December;13(12):490-4.

(24) Madi A, Kenett DY, Bransburg-Zabary S, Merbl Y, Quintana FJ, Tauber AI et al. Network theory analysis of antibody-antigen reactivity data: the immune trees at birth and adulthood. PLoS ONE 2011;6(3):e17445.

(25) Cohen IR. Autoantibody repertoires, natural biomarkers, and system controllers. Trends Immunol 2013 December;34(12):620-5.

(26) Silverman GJ, Gronwall C, Vas J, Chen Y. Natural autoantibodies to apoptotic cell membranes regulate fundamental innate immune functions and suppress inflammation. Discov Med 2009 October;8(42):151-6.

(27) Sachtleben P, Schmidt WA, Klein G. Die elektrokinetischen Potentiale von Gewebekulturzellen nach Infektion mit Coxsackie-B3 Virus [The electrokinetic potentials of tissue culture cells after infection with coxsackie B3-Virus]. Arch Gesamte Virusforsch 1967;20(1):99-108.

(28) Thompson CJ, Docherty JJ, Boltz RC, Gaines RA, Todd P. Electrokinetic alteration of the surface of herpes simplex virus infected cells. J Gen Virol 1978 June;39(3):449-61.

(29) Mehrishi JN, Bauer J. Electrophoresis of cells and the biological relevance of surface charge. Electrophoresis 2002 July;23(13):1984-94.

(30) Papanikolaou E, Paruzynski A, Kasampalidis I, Deichmann A, Stamateris E, Schmidt M et al. Cell cycle status of CD34(+) hemopoietic stem cells determines lentiviral integration in actively transcribed and development-related genes. Mol Ther 2015 April;23(4):683-96.

(31) Forterre P, Prangishvili D. The great billion-year war between ribosome- and capsid-encoding organisms (cells and viruses) as the major source of evolutionary novelties. Ann N Y Acad Sci 2009 October;1178:65-77.

(32) Villarreal LP. The source of self: genetic parasites and the origin of adaptive immunity. Ann N Y Acad Sci 2009 October;1178:194-232.

(33) Villarreal LP. Persistence pays: how viruses promote host group survival. Curr Opin Microbiol 2009 August;12(4):467-72.

(34) Villarreal LP, Witzany G. Viruses are essential agents within the roots and stem of the tree of life. J Theor Biol 2010 February 21;262(4):698-710.

(35) Villarreal LP. Viral ancestors of antiviral systems. Viruses 2011 October;3(10):1933-58.

(36) Forterre P, Prangishvili D. The major role of viruses in cellular evolution: facts and hypotheses. Curr Opin Virol 2013 October;3(5):558-65.

(37) Crisp A, Boschetti C, Perry M, Tunnacliffe A, Micklem G. Expression of multiple horizontally acquired genes is a hallmark of both vertebrate and invertebrate genomes. Genome Biology 2015;16(1):50.

(38) Krämer A, Kahn MH. Global Challenges of Infectious Disease Epidemiology. In: Krämer et al., editor. Modern Infectious Disease Epidemiology. Springer; 2010. p. 23.

(39) Mukherjee S. A Private Plague. The Emperor of All Maladies; A biography of cancer. 1 ed. London: Fourth Estate; 2011. p. 37-44.

(40) Klein G. Toward a genetics of cancer resistance. PNAS 2009 January 20;106(3):859-63.

(41) Ondov BD, Bergman NH, Phillippy AM. Interactive metagenomic visualization in a Web browser. BMC Bioinformatics 2011;12:385.

(42) Friberg S, Mattson S. On the growth rates of human malignant tumors: implications for medical decision making. J Surg Oncol 1997 August;65(4):284-97.

(43) de Magalhaes JP. How ageing processes influence cancer. Nat Rev Cancer 2013 May;13(5):357-65.

(44) Beerenwinkel N, Schwarz RF, Gerstung M, Markowetz F. Cancer evolution: mathematical models and computational inference. Syst Biol 2015 January;64(1):e1-25.

(45) Pardoll D. Does the immune system see tumors as foreign or self? Annu Rev Immunol 2003;21:807-39.

(46) Pradeu T, Jaeger S, Vivier E. The speed of change: towards a discontinuity theory of immunity? Nat Rev Immunol 2013 October;13(10):764-9.

(47) Alberts B, Johnson A, Lewis J, Raff M, Roberts K, Walter P. **Molecular Biology of the Cell.** 4 ed. New York : Garland Publishing; 2002.

(48) Udagawa T. Tumor dormancy of primary and secondary cancers. APMIS 2008 July;116(7-8):615-28.

(49) Black WC, Welch HG. Advances in diagnostic imaging and overestimations of disease prevalence and the benefits of therapy. N Engl J Med 1993 April 29;328(17):1237-43.

(50) Martincorena I, Campbell PJ. Somatic mutation in cancer and normal cells. Science 2015 September 25;349(6255):1483-9.

(51) Marches R, Scheuermann R, Uhr J. Cancer dormancy: from mice to man. Cell Cycle 2006 August;5(16):1772-8.

(52) Guidotti LG, Chisari FV. To kill or to cure: options in host defense against viral infection. Curr Opin Immunol 1996 August;8(4):478-83.

(53) Thimme R, Oldach D, Chang KM, Steiger C, Ray SC, Chisari FV. Determinants of viral clearance and persistence during acute hepatitis C virus infection. J Exp Med 2001 November 19;194(10):1395-406.

(54) Thompson AE. JAMA patient page. The immune system. JAMA 2015 April 28;313(16):1686.

(55) Bretscher P, Cohn M. A theory of self-nonself discrimination. Science 1970 September 11;169(950):1042-9.

(56) Germain RN. The art of the probable: system control in the adaptive immune system. Science 2001 July 13;293(5528):240-5.

(57) Gett AV, Hodgkin PD. A cellular calculus for signal integration by T cells. Nat Immunol 2000 September;1(3):239-44.

(58) Marchingo JM, Kan A, Sutherland RM, Duffy KR, Wellard CJ, Belz GT et al. T cell signaling. Antigen affinity, costimulation, and cytokine inputs sum linearly to amplify T cell expansion. Science 2014 November 28;346(6213):1123-7.

(59) Mehrishi JN, Szabo M, Bakacs T. Some aspects of the recombinantly expressed humanised superagonist anti-CD28 mAb, TGN1412 trial catastrophe lessons to safeguard mAbs and vaccine trials. Vaccine 2007 May 4;25(18):3517-23.

(60) Suntharalingam G, Perry MR, Ward S, Brett SJ, Castello-Cortes A, Brunner MD et al. Cytokine storm in a phase 1 trial of the anti-CD28 monoclonal antibody TGN1412. N Engl J Med 2006 September 7;355(10):1018-28.

(61) Vitetta ES, Ghetie VF. Immunology. Considering therapeutic antibodies. Science 2006 July 21;313(5785):308-9.

(62) Weinberg RA. Robert Weinberg: Beyond Hallmarks. Trends in Cancer 2015 September;1(1):4-5.

(63) Gubin MM, Schreiber RD. CANCER. The odds of immunotherapy success. Science 2015 October 9;350(6257):158-9.

(64) Allison JP. Immune Checkpoint Blockade in Cancer Therapy: The 2015 Lasker-DeBakey Clinical Medical Research Award. JAMA 2015 September 8;1113-5.

(65) Postow MA, Callahan MK, Wolchok JD. Immune Checkpoint Blockade in Cancer Therapy. J Clin Oncol 2015 June 10;33(17):1974-82.

(66) Bobbio A, Alifano M. Immune therapy of non-small cell lung cancer. The future. Pharmacol Res 2015 September;99:217-22.

(67) Pawankar R, Canonica WG, Holgate ST, Lockey ST. Introduction and Executive Summary. In: Pawankar R, Canonica WG, Holgate, S.T., Lockey RF, editors. White Book on Allergy.Milwaukee, Wisconsin: World Allergy Organization (WAO); 2011. p. 11-20.

(68) Umetsu DT, McIntire JJ, Akbari O, Macaubas C, DeKruyff RH. Asthma: an epidemic of dysregulated immunity. Nat Immunol 2002 August;3(8):715-20.

(69) Timp W, Feinberg AP. Cancer as a dysregulated epigenome allowing cellular growth advantage at the expense of the host. Nat Rev Cancer 2013 July;13(7):497-510.

(70) Liu X, Jiang S, Fang C, Yang S, Olalere D, Pequignot EC et al. Affinity-Tuned ErbB2 or EGFR Chimeric Antigen Receptor T Cells Exhibit an Increased Therapeutic Index against Tumors in Mice. Cancer Res 2015 September 1;75(17):3596-607.

(71) Szabados T, Tusnady G, Varga L, Bakacs T. A stochastic model of b cell affinity maturation and a network model of immune memory. arXiv.org . 2015. Ref Type: Internet Communication

(72) Benoist C, Mathis D. Autoimmunity provoked by infection: how good is the case for T cell epitope mimicry? Nat Immunol 2001 September;2(9):797-801.